\documentclass[runningheads]{llncs}
\usepackage[T1]{fontenc}
\usepackage{graphicx}
\begin{document}
\title{Assessing the Influence of Different Types of Probing on Adversarial Decision-Making in a Deception Game}
\author{Md Abu Sayed\inst{1} \and Mohammad Ariful Islam Khan \inst{1}   \and Bryant A Allsup\inst{1} \and 
Joshua Zamora\inst{1} \and Palvi Aggarwal\inst{1}}
\authorrunning{M. A. Sayed et al.}
\institute{University of Texas at El Paso, TX 79968, USA \\
\email{msayed@miners.utep.edu, mkhan16@miners.utep.edu, baallsup@miners.utep.edu, jazamora6@miners.utep.edu, paggarwal@utep.edu}}
\maketitle              % typeset the header of the contribution
\begin{abstract}
Deception, which includes leading cyber-attackers astray with false information, has shown to be an effective method of thwarting cyber-attacks. There has been little investigation of the effect of probing action costs on adversarial decision-making, despite earlier studies on deception in cybersecurity focusing primarily on variables like network size and the percentage of honeypots utilized in games. Understanding human decision-making when prompted with choices of various costs is essential in many areas such as in cyber security. In this paper, we will use a deception game (DG) to examine different costs of probing on adversarial decisions. To achieve this we utilized an IBLT model and a delayed feedback mechanism to mimic knowledge of human actions. Our results were taken from an even split of deception and no deception to compare each influence. It was concluded that probing was slightly taken less as the cost of probing increased. The proportion of attacks stayed relatively the same as the cost of probing increased. Although a constant cost led to a slight decrease in attacks. Overall, our results concluded that the different probing costs do not have an impact on the proportion of attacks whereas it had a slightly noticeable impact on the proportion of probing.

\end{abstract}
\section{Introduction}
Cyber attacks are increasing exponentially day by day as the digital space becomes more prevalent. These attacks involve unwelcome attempts to steal, expose, alter, disable or destroy information through unauthorized access to computer systems \cite{comptiasecurity}. Current solutions to combat these cyber attacks have proven somewhat sufficient such as intrusion detection systems (IDSs) \cite{mell2003overview,shang2018false} and filtering strategies \cite{shang2018hybrid}. Intrusion detection systems operate by searching for traces of known attacks or deviations from normal activity whereas filtering strategies filter out activities that may be seen as malicious . Although, the need for increasingly dependable infrustructures to combat these attacks are in demand now more than ever. 

Therefore, the use of deception has proven advantageous as an attempt to combat these attacks \cite{aggarwal2016cyber}. Deception involves providing false information in order to deceive a person into believing said false information. Deception is used in cyber security as a technique where honeypots act as decoys to prevent actual web servers from being attacked \cite{almeshekah2016cyber,aggarwal2016looking,islam2019performance,khan2021iot,sayed2022cyber,abu2023cyber,sayed2023honeypot}. 

Machine learning is essential for improving cognitive modeling because it makes data-driven methods and insights possible. Research has looked at a number of topics, such as utilizing machine learning to direct cognitive modeling, deep learning to predict human judgments, adding human cognitive biases into machine learning models, and improving cognitive models by combining knowledge from human memory models \cite{agrawal2019using,sayed2017understanding,taniguchi2018machine,emon2020automatic,haque2021covid,raju2020predicting,emon2020automated,mahmud2023machine,fazle2023novel}.

There have been various methods to help understand the human nature of these attacks which have been by the use of Instance Based Learning (IBL). This model will be used since it may provide insight into human decision making in adverse scenarios involving cyber attacks \cite{gonzalez2011instance,gonzalez2003instance}.

Furthermore, probing is an under utilized factor that must be considered. For instance, when adversaries make a decision on which server to attack they often probe before making an attack which will provide information about the server beforehand. To realize the effect of adversary decisions, the cost of these probing actions is the primary focus of this paper.

We developed a deception game to understand the effect of these probing actions. We believe that understanding these probing action costs will provide insight into adversary actions and as a result this will help better predict and combat cyber attacks.

The rest of this paper is organized as follows. Section \ref{sec:td} discusses the task description. Section \ref{sec:mt} describes methodology,  and section \ref{sec:re} discusses result. Finally, conclusions  are presented in Section \ref{sec:conc}.

\section{Task Description}\label{sec:td}

The deception game(DG) is a single user multi-choice game where the goal is to figure out the correct server to attack \cite{garg2007deception}.  In this game Users are tasked with attacking one real server when presented with four servers, two real and two fake(honeypots), over the course of 30 rounds and at the end of each round users will be given feedback(Delayed Feedback). Users are given signals by the system to tell users if they are honeypots or if they are real. Through 15 of these thirty rounds the server will lie about what is being signaled \cite{nguyen2020effects}.

When the game begins users will take the role of the attacker and the system will play defender. The first thing that will happen in the game is the user will decide which server to probe for information, users can probe as many servers as they want. This is known as the probing stage and is where the signals will appear to notify the user if the attacker is attacking a honeypot or a real server. It will be up to the user to determine here if this is a deception round(user is lied to) or if this is the truth. 

Once the user chooses they will then enter the attack stage. Here users are looking at the one server they chose and are given two options, attack or retreat. If the user attacks, the round will be over and the delayed feed-back will be given telling the user if they won or lost points for the round. If the user retreats the round will be over and the results will not go for or against the user. (need to know if the feedback is given, revealing if it was a honey pot or not). Once the 30 rounds are done with all the participants we will use the results to calculate the score.
	
The score is calculated with three different metrics in mind. There is the no cost, set cost and incremental cost. A baseline metric that is true across all the metrics will be attacking a correct server will give you 10 points, attacking a honeypot will lose you 10 points and not attacking will give 0 points. The first metric we will use is the no cost metric. The no cost will be the baseline and will only calculate points based on decisions made in the attacking stage. The next metric is set cost. 

In the set cost metric users are given points based on if they probed the correct server, so if a user probes a server and it ends up being a honeypot then the user will be given -5 points and if the user probes a server and it is a real server the user is given +5 points. This info is still withheld until the delayed feedback is given as the score is based on the servers actual property and not the one that is shown. The final scoring system is the incremental. The incremental will resemble the set cost with the difference being instead of +5 or -5 points when probing a server the scoring is changed to 5(number of real servers probed prior) and -5*h(number of honeypot servers that you probed prior).  

\begin{table}[!htb]
    %\captionsetup{justification=centering}
    \caption{Deception Cost}
    \label{table:1}
    \centering
    \small
    \begin{tabular} {|c|c|c|c|} \hline
        Server/ Cost  & No Cost & Set Cost & Incremental Cost \\ \hline
Real & 0 & 5 & 5 \\ \hline
Honeypot & 0 & -5 & -5*h \\ \hline
No Probing & 0 & 0 & 0 \\ \hline
    \end{tabular}
\end{table}

\begin{table}[!htb]
    %\captionsetup{justification=centering}
    \caption{Attack Cost}
    \label{table:1}
    \centering
    \small
    \begin{tabular} {|c|c|c|c|} \hline
        Server/ Cost  & No Cost & Set Cost & Incremental Cost \\ \hline
Real & 10 & 10 & 10 \\ \hline
Honeypot & -10 & -10 & -10 \\ \hline
    \end{tabular}
\end{table}

In this experiment we will be using the IBLT Model to replicate human experience found in \cite{katakwar2022influence}. We decided to use the IBLT model as we were familiar with it and it allows us to replicate results to found in the paper \cite{katakwar2022influence}. In order to replicate these results our team also implemented items such as delayed feedback.

Delayed feedback is a method that gives a response to a user at the end of a round. The prior method is that as soon as a decision is made the feedback is given. The problem with this is that responses are made within the activity and allows for changes to be made when it happens, tampering with each instance of a problem. This would primarily be seen in the probing stage of the set and incremental point metrics where a user would receive or lose points in the probing stage. Since we changed to delayed feedback no additional information would be given that would help agents during each round. This allows us to better use the different points metric and capture a more accurate representation \cite{nguyen2020effects}. 

The different costs used in the deception game were used to replicate the study. This is the same for the number of trials given. The number of participants is an arbitrary number to get a wider look if the trends stayed consistent in the human experiment. 

\section{Methodology}\label{sec:mt}

\subsection{Experimental Design}

In our pyibl modling, initially we strat with 40 participant in  no-cost probe, constant-cost probe, and increasing-cost probe  and 30 trial for each participant. Each of these conditions had the same number of webservers, and the proportion of honeypot webservers in the network was constant across all the conditions (50\%). We keep four webservers, out of the four webservers, 2 of them were honeypot webservers, while the other 2 were regular webservers. In no-cost condition, the player had no cost for their probe actions on honeypot webservers in the probe stage of DG. In constant-cost conditions, the honeypot webserver probe cost remained constant(5) with the repeated probe attempts on honeypots. In increasing cost conditions, the honeypot webserver probe's cost increased linearly(5*h, number of interaction with honeypot server) with the repeated probe attempts to honeypot servers. Around all the conditions, each of the rounds had a probe stage followed by the attack stage.

Each of the three conditions had 30 trials, out of which 15 of them were randomly assigned as deception rounds, while the rest of the rounds had no deception. Also, the deception and non-deception rounds in DG did not form a particular sequence or pattern, that’s why it’s to predict.

\subsection{Deception Game}

\begin{figure}[h!]
    \centering
    \includegraphics[width=10cm, height=6cm]{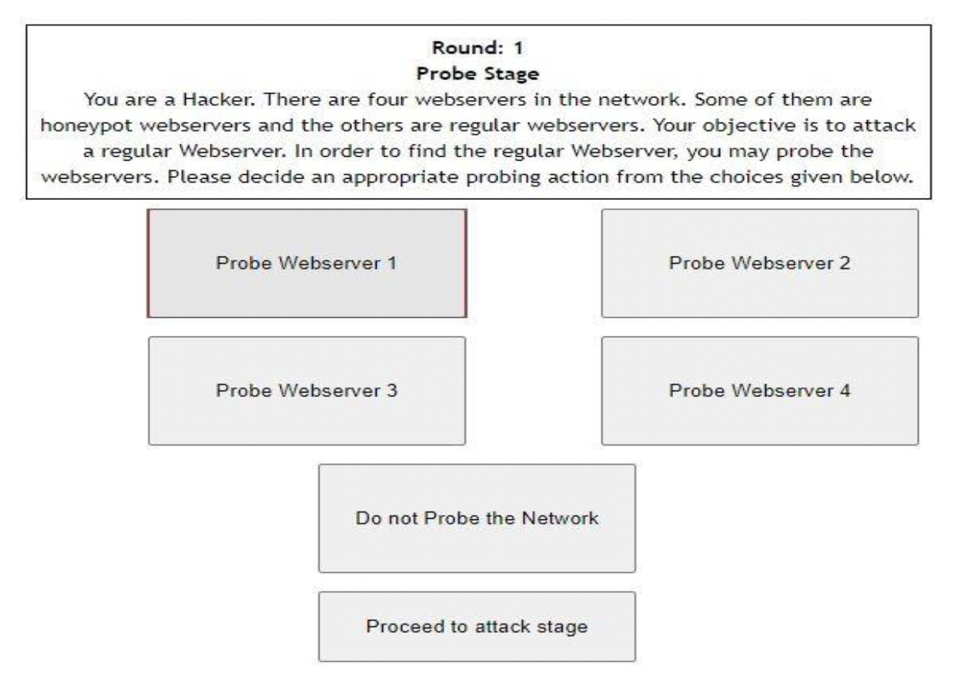}
    \caption{Probe stage of Deception Game in increasing cost condition \cite{katakwar2022influence}.}
    \label{fig:fig1}
\end{figure}

\begin{figure}[h!]
    \centering
    \includegraphics[width=10cm, height=6cm]{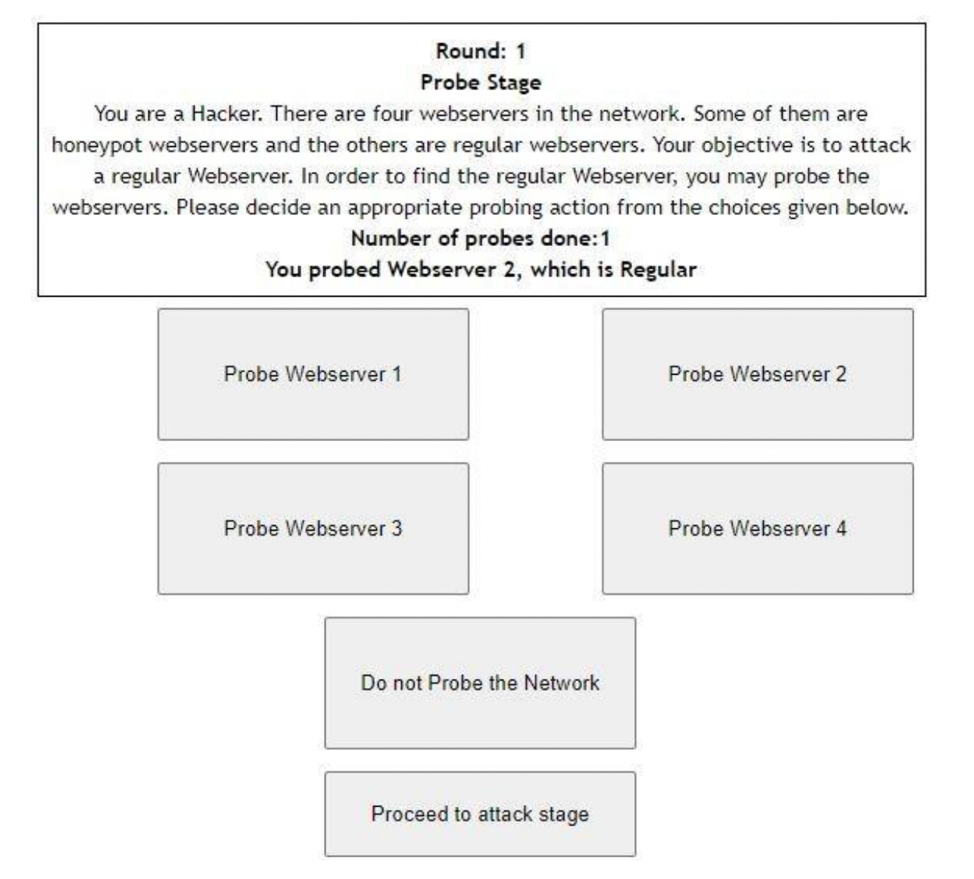}
    \caption{Response received form system at probe stage \cite{katakwar2022influence}.}
    \label{fig:fig2}
\end{figure}

\begin{figure}[h!]
    \centering
    \includegraphics[width=10cm, height=6cm]{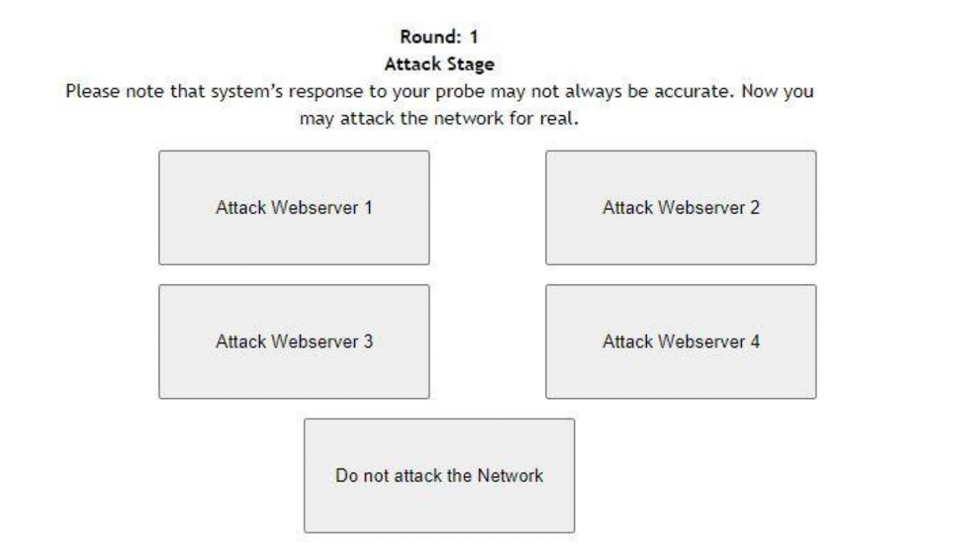}
    \caption{Attack stage \cite{katakwar2022influence}.}
    \label{fig:fig3}
\end{figure}

\begin{figure}[h!]
    \centering
    \includegraphics[width=10cm, height=6cm]{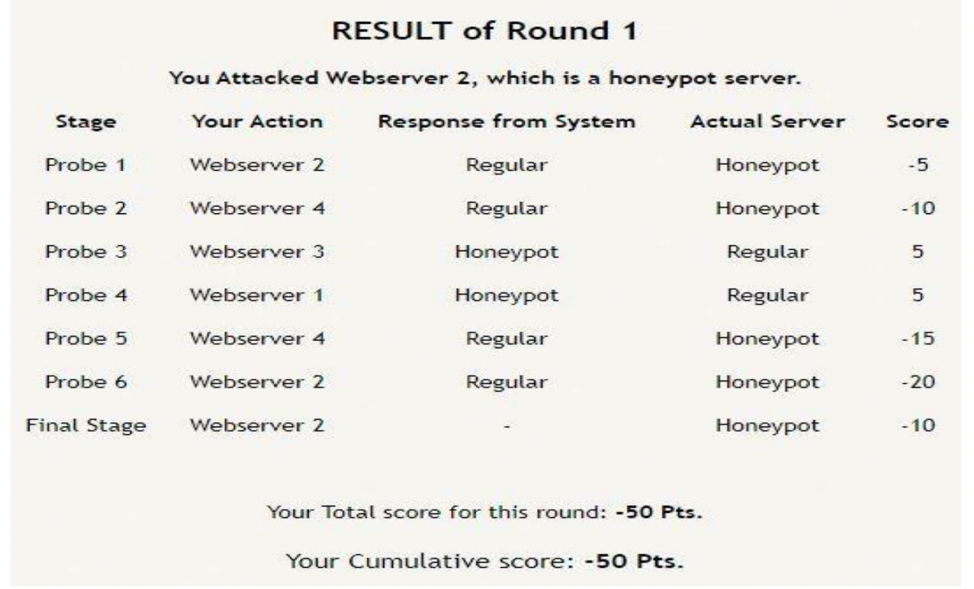}
    \caption{Complete result of the round for increasing cost \cite{katakwar2022influence}.}
    \label{fig:fig4}
\end{figure}

We are showing some screenshots of our developed deception  game model only for increasing cost. Fig. ~\ref{fig:fig1} demonstrates the initial interface in a particular round form where player can probe any webserver and then go to attack stage or dirrectly go to the attack stage . Response(signal, not outcome) received after selecting a particular webserver has shown in fig. ~\ref{fig:fig2}. Fig. ~\ref{fig:fig3} depicts how attack attack looks like which is more or less like probe stage. Finally, fig. ~\ref{fig:fig4} showed one full round of increasing cost condition.

\section{Results}\label{sec:re}

\subsection{Influence of different cost conditions over the trials on adversarial decisions during probe stage}

\begin{figure}[h!]
    \centering
    \includegraphics[width=10cm, height=6cm]{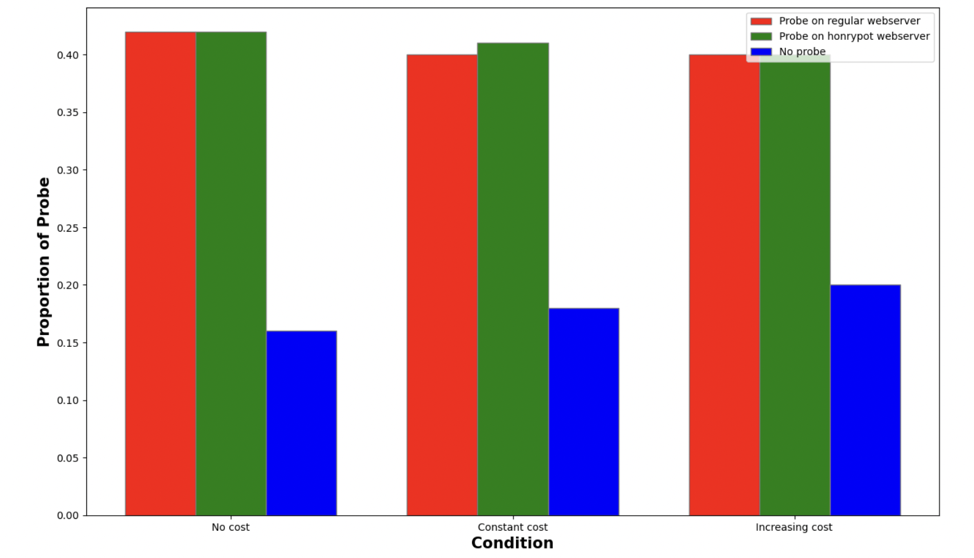}
    \caption{Impact different probing-cost at probe stage.}
    \label{fig:fig5}
\end{figure}

We observe the effect of different cost condition on adversarial decision making in probe stage. We saw probing percentage on regular webserver is almost same over three cost condition (0.42, 0.40, 0.40). Probing percentage on honeypot webserver also followed the same pattern (0.42, 0.41, 0.40) showed in fig. ~\ref{fig:fig5}.

Interestingly, no probing percentage slightly increase(0.16, 0.18, 0.20) over the three cost conditions (no cost, constant cost, increasing cost) . The main reason behind that when cost increases, agent will choose no probe because of no associated cost. The percentage data of probe on regular, honeypot webserver, we collect  from our pyibl modeling. Overall, form figure, we see different probling cost conditions over the trial do not have any impact on the adversarial decision making at probe stage.

\subsection{Influence of different cost conditions over the trials on adversarial decisions during attack stage}

\begin{figure}[h!]
    \centering
    \includegraphics[width=10cm, height=6cm]{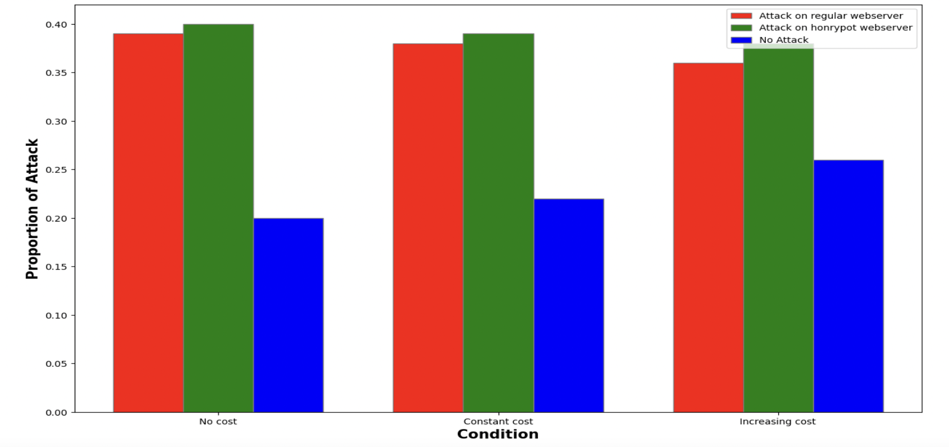}
    \caption{Impact different probing-cost at attack stage}
    \label{fig:fig6}
\end{figure}

In every trial, attack decision takes place only once, whereas probing stage happens at most 5 times. We also observe the effect of different cost condition on adversarial decision making in attack stage. Attack  percentage on regular webserver is almost same over three cost condition(0.39, 0.38, 0.36) . Percentage of attack on honeypot webserver also followed the same pattern (0.40, 0.39, 0.38) showed in Fig. ~\ref{fig:fig6}.

It is worth noting that, no attack percentage slightly increase(0.20, 0.22, 0.26) over the three cost conditions (no cost, constant cost, increasing cost). The percentage data of attack on regular, honeypot webserver, we collect  from our pyibl modeling. Overall, form figure, we see different probling cost conditions over the trial do not have any impact on the adversarial decision making at attack stage. Interestingly, when we set pre-population value to a small value, we do not see specific pattern, then we slightly increase the pre-population and see the pattern.

\subsection{Analysis of human data}

\begin{figure}[h!]
    \centering
    \includegraphics[width=10cm, height=6cm]{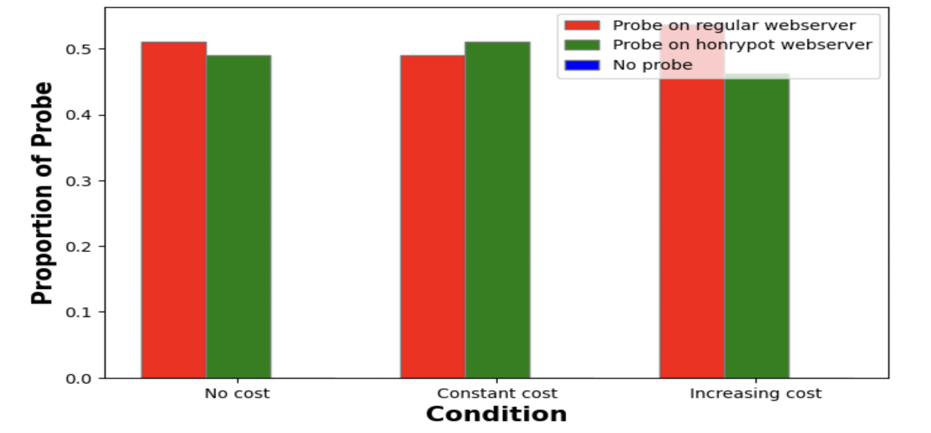}
    \caption{Impact different probing-cost at probe stage(human data)}
    \label{fig:fig7}
\end{figure}

\begin{figure}[h!]
    \centering
    \includegraphics[width=10cm, height=6cm]{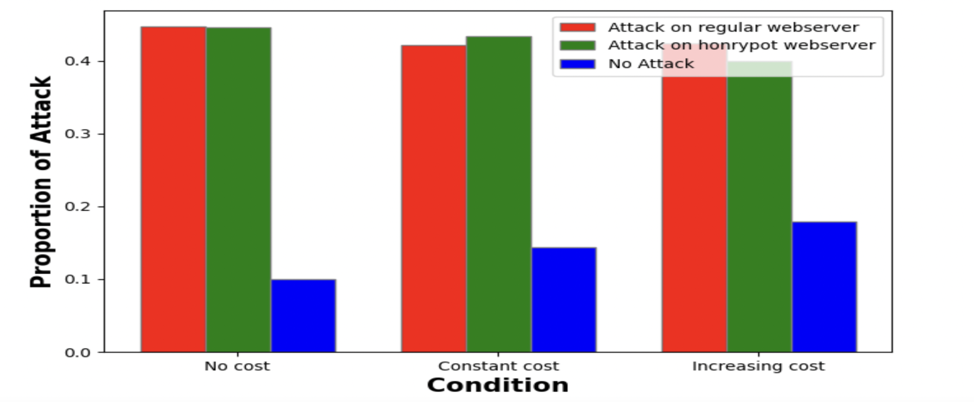}
    \caption{Impact different probing-cost at attack stage(human data)}
    \label{fig:fig8}
\end{figure}

We also collected human data from work \cite{katakwar2022influence} we trying to replicating. Fig. ~\ref{fig:fig7} and ~\ref{fig:fig8} shows proportion of probe and attack on different cost conditions. Interestingly in human data, no probe never happened. In fig. ~\ref{fig:fig7}, we see probability of probe stays almost same on different cost conditions. In fig. ~\ref{fig:fig8}, we see probability of attack also same accross different cost conditions.

\subsection{Compare with previous work}

\begin{figure}[h!]
    \centering
    \includegraphics[width=10cm, height=6cm]{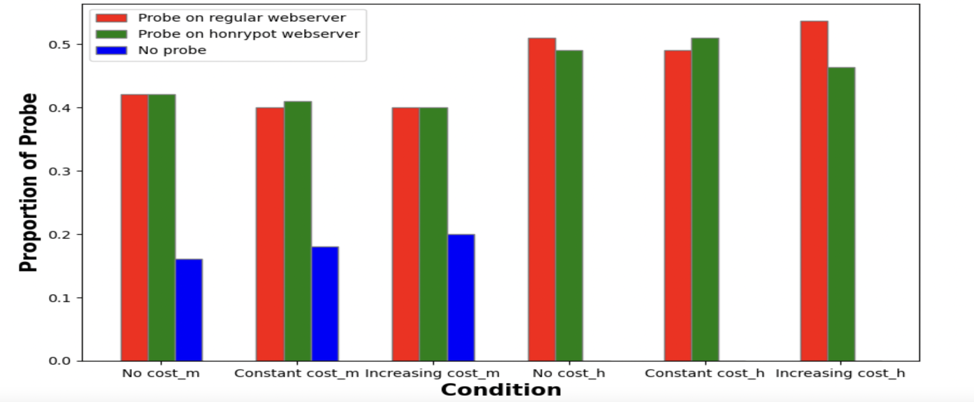}
    \caption{Comparison of result in probe stage}
    \label{fig:fig9}
\end{figure}

\begin{figure}[h!]
    \centering
    \includegraphics[width=10cm, height=6cm]{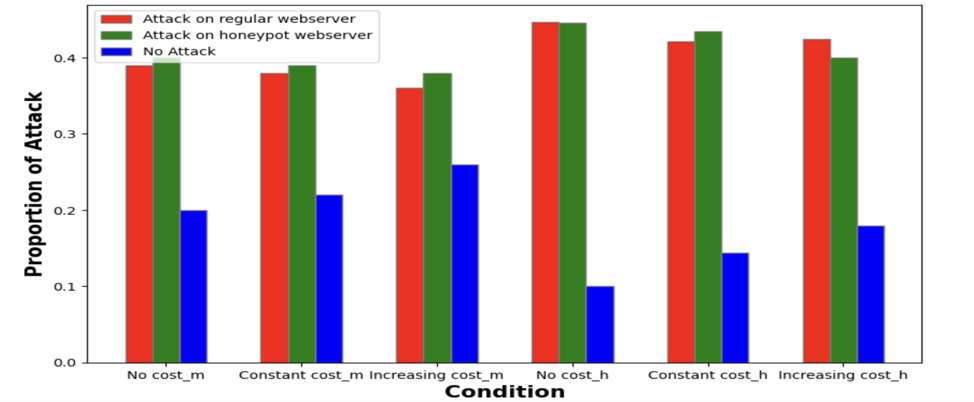}
    \caption{Comparison of result in attack stage}
    \label{fig:fig10}
\end{figure}

We also compare our finding with human data. In our modeling, we fixed the number of probing by 5, but in human data number of is not fixed. That’s why direct comparison with our model result and human data result is not possible. Therefore, we are trying to follow common patterns. Fig. ~\ref{fig:fig9} and ~\ref{fig:fig10} compares our finding with human data. In probe stage (Fig. ~\ref{fig:fig9}), we see both our model and human data follows same pattern, only in human data no values for no probe conditions. In attack stage (Fig. ~\ref{fig:fig10}), different probing costs do not have significant impact on attack decision in both modeling and human data, also here no attack follows increasing trend in both cases. 

\section{Conclusion}\label{sec:conc}

Using honeypots as a deception technique has demonstrated it’s vitality as a tactic for thwarting current cyberattacks. Non-canonical games have been designed and utilized by researchers in the field of adversarial cybersecurity to investigate the efficacy of deception in various cybersecurity scenarios. People have also looked into the numerous human aspects that influence the attacker's judgment in deception-based games.

Researchers have previously investigated the influence of network size on adversarial decisions in a deception game \cite{katakwar2022influence}. Moreover, Gonzalez et al. have worked on the influence of probing action costs on adversarial decision-making in such games \cite{gonzalez2021sequential}. This work involves lab-based experiment to understand the impact of probing action costs on adversarial decisions. However, there is a lack of work regarding the IBL modeling of the attacker decision-making in a deception game. In our work, we have addressed this issue and developed an IBL model which helps us to better understand the attacker actions in cybersecurity situations. Our results demonstrate certain significant implications for real-world cyber attacks scenarios. If we consider different conditions like no cost, constant cost and increasing cost it is quite conspicuous that the attack-decision is identical in those situations. Hence, it can be said that probing action cost has no impact on adversary-decsions. In other words, making probes on the network expensive to attackers might not influence their attack decisions which conforms with the previous work \cite{gonzalez2021sequential}.

Since there is a usage of real-world cyberattack variables in DG, where the attacker (or the model) first investigates the state of the webservers before launching an assault, some of the findings from our experiment could be applicable in the real-life situations. In the future, we intend to work on the sequential decisions from sampling in a cyber-attack scenario \cite{gonzalez2021sequential}.

\bibliographystyle{unsrt}
\bibliography{reference}

\begin{thebibliography}{10}

\bibitem{comptiasecurity}
CompTIA.
\newblock Network security: What is it, why does it matter and what can you do
  to make networks more secure?
\newblock
  \url{https://www.comptia.org/content/guides/network-security-basics-definition-threats-and-solutions}.
\newblock Accessed: Mar, 2023.

\bibitem{mell2003overview}
Peter Mell, Vincent Hu, Richard Lippmann, Josh Haines, and Marc Zissman.
\newblock An overview of issues in testing intrusion detection systems.
\newblock 2003.

\bibitem{shang2018false}
Yilun Shang.
\newblock False positive and false negative effects on network attacks.
\newblock {\em Journal of Statistical Physics}, 170(1):141--164, 2018.

\bibitem{shang2018hybrid}
Yilun Shang.
\newblock Hybrid consensus for averager--copier--voter networks with
  non-rational agents.
\newblock {\em Chaos, Solitons \& Fractals}, 110:244--251, 2018.

\bibitem{aggarwal2016cyber}
Palvi Aggarwal, Cleotilde Gonzalez, and Varun Dutt.
\newblock Cyber-security: role of deception in cyber-attack detection.
\newblock In {\em Advances in Human Factors in Cybersecurity: Proceedings of
  the AHFE 2016 International Conference on Human Factors in Cybersecurity,
  July 27-31, 2016, Walt Disney World{\textregistered}, Florida, USA}, pages
  85--96. Springer, 2016.

\bibitem{almeshekah2016cyber}
Mohammed~H Almeshekah and Eugene~H Spafford.
\newblock Cyber security deception.
\newblock {\em Cyber Deception: Building the Scientific Foundation}, pages
  23--50, 2016.

\bibitem{aggarwal2016looking}
Palvi Aggarwal, Cleotilde Gonzalez, and Varun Dutt.
\newblock Looking from the hacker's perspective: Role of deceptive strategies
  in cyber security.
\newblock In {\em 2016 International conference on cyber situational awareness,
  data analytics and assessment (CyberSA)}, pages 1--6. IEEE, 2016.

\bibitem{islam2019performance}
Saiful Islam, Md~Ariful~Islam Khan, Sanjida~Tasnim Shorno, Sumonto Sarker, and
  Md~Abubakar Siddik.
\newblock Performance evaluation of sdn controllers in wireless network.
\newblock In {\em 2019 1st International Conference on Advances in Science,
  Engineering and Robotics Technology (ICASERT)}, pages 1--5. IEEE, 2019.

\bibitem{khan2021iot}
Mohammad Ariful~Islam Khan, Mohammad~Akidul Hoque, and Sabbir Ahmed.
\newblock Iot-based system for real-time water pollution monitoring of rivers.
\newblock In {\em 2021 International Conference on Electronics, Communications
  and Information Technology (ICECIT)}, pages 1--5. IEEE, 2021.

\bibitem{sayed2022cyber}
Md~Abu Sayed, Ahmed~H Anwar, Christopher Kiekintveld, Branislav Bosansky, and
  Charles Kamhoua.
\newblock Cyber deception against zero-day attacks: a game theoretic approach.
\newblock In {\em International Conference on Decision and Game Theory for
  Security}, pages 44--63. Springer, 2022.

\bibitem{abu2023cyber}
Md~Abu~Sayed, Ahmed~H Anwar, Christopher Kiekintveld, Branislav Bosansky, and
  Charles Kamhoua.
\newblock Cyber deception against zero-day attacks: A game theoretic approach.
\newblock {\em arXiv e-prints}, pages arXiv--2307, 2023.

\bibitem{sayed2023honeypot}
Md~Abu Sayed, Ahmed~H Anwar, Christopher Kiekintveld, and Charles Kamhoua.
\newblock Honeypot allocation for cyber deception in dynamic tactical networks:
  A game theoretic approach.
\newblock {\em arXiv preprint arXiv:2308.11817}, 2023.

\bibitem{agrawal2019using}
Mayank Agrawal, Joshua~C Peterson, and Thomas~L Griffiths.
\newblock Using machine learning to guide cognitive modeling: A case study in
  moral reasoning.
\newblock {\em arXiv preprint arXiv:1902.06744}, 2019.

\bibitem{sayed2017understanding}
Md~Abu Sayed, Md~Maksudur Rahman, Moinul~Islam Zaber, and Amin~Ahsan Ali.
\newblock Understanding dhaka city traffic intensity and traffic expansion
  using gravity model.
\newblock In {\em 2017 20th International Conference of Computer and
  Information Technology (ICCIT)}, pages 1--6. IEEE, 2017.

\bibitem{taniguchi2018machine}
Hidetaka Taniguchi, Hiroshi Sato, and Tomohiro Shirakawa.
\newblock A machine learning model with human cognitive biases capable of
  learning from small and biased datasets.
\newblock {\em Scientific reports}, 8(1):7397, 2018.

\bibitem{emon2020automatic}
Solayman~Hossain Emon, AHM Annur, Abir~Hossain Xian, Kazi~Mahia Sultana, and
  Shoeb~Mohammad Shahriar.
\newblock Automatic video summarization from cricket videos using deep
  learning.
\newblock In {\em 2020 23rd International Conference on Computer and
  Information Technology (ICCIT)}, pages 1--6. IEEE, 2020.

\bibitem{haque2021covid}
Samiul Haque, Mohammad~Akidul Hoque, Mohammad Ariful~Islam Khan, and Sabbir
  Ahmed.
\newblock Covid-19 detection using feature extraction and semi-supervised
  learning from chest x-ray images.
\newblock In {\em 2021 IEEE Region 10 Symposium (TENSYMP)}, pages 1--5. IEEE,
  2021.

\bibitem{raju2020predicting}
Muntaqim~Ahmed Raju, Md~Solaiman Mia, Md~Abu Sayed, and Md~Riaz Uddin.
\newblock Predicting the outcome of english premier league matches using
  machine learning.
\newblock In {\em 2020 2nd International Conference on Sustainable Technologies
  for Industry 4.0 (STI)}, pages 1--6. IEEE, 2020.

\bibitem{emon2020automated}
Solayman~Hossain Emon, MD~Afranul~Haque Mridha, and Mohtasim Shovon.
\newblock Automated recognition of rice grain diseases using deep learning.
\newblock In {\em 2020 11th International Conference on Electrical and Computer
  Engineering (ICECE)}, pages 230--233. IEEE, 2020.

\bibitem{mahmud2023machine}
Sultan Mahmud, Md~Mohsin, Abdul Muyeed, Shaila Nazneen, Md~Abu Sayed, Nabil
  Murshed, Tajrin~Tahrin Tonmon, and Ariful Islam.
\newblock Machine learning approaches for predicting suicidal behaviors among
  university students in bangladesh during the covid-19 pandemic: A
  cross-sectional study.
\newblock {\em Medicine}, 102(28), 2023.

\bibitem{fazle2023novel}
Md~Fazle~Rabbi, Solayman Hossain~Emon, Ehtesham Mahmud~Nishat, Atira Ferdoushi,
  Chun-Che Huang, Md~Fashiar~Rahman, et~al.
\newblock A novel approach for defect detection of wind turbine blade using
  virtual reality and deep learning.
\newblock {\em arXiv e-prints}, pages arXiv--2401, 2023.

\bibitem{gonzalez2011instance}
Cleotilde Gonzalez and Varun Dutt.
\newblock Instance-based learning: integrating sampling and repeated decisions
  from experience.
\newblock {\em Psychological review}, 118(4):523, 2011.

\bibitem{gonzalez2003instance}
Cleotilde Gonzalez, Javier~F Lerch, and Christian Lebiere.
\newblock Instance-based learning in dynamic decision making.
\newblock {\em Cognitive Science}, 27(4):591--635, 2003.

\bibitem{garg2007deception}
Nandan Garg and Daniel Grosu.
\newblock Deception in honeynets: A game-theoretic analysis.
\newblock In {\em 2007 IEEE SMC information assurance and security workshop},
  pages 107--113. IEEE, 2007.

\bibitem{nguyen2020effects}
Thuy~Ngoc Nguyen and Cleotilde Gonzalez.
\newblock Effects of decision complexity in goal-seeking gridworlds: A
  comparison of instance-based learning and reinforcement learning agents.
\newblock In {\em Proceedings of the 18th intl. conf. on cognitive modelling},
  2020.

\bibitem{katakwar2022influence}
Harsh Katakwar, Palvi Aggarwal, Zahid Maqbool, and Varun Dutt.
\newblock Influence of probing action costs on adversarial decision-making in a
  deception game.
\newblock In {\em Ict analysis and applications}, pages 649--658. Springer,
  2022.

\bibitem{gonzalez2021sequential}
Cleotilde Gonzalez and Palvi Aggarwal.
\newblock Sequential decisions from sampling: Inductive generation of stopping
  decisions using instance-based learning theory.
\newblock 2021.

\end{thebibliography}
\end{document}